\documentclass[journal=jacsat,manuscript=article]{achemso}

\usepackage[version=3]{mhchem} 
\usepackage{xcolor}
\usepackage{amsmath}
\usepackage{siunitx}
\usepackage{color,soul}
\usepackage{overpic}
\usepackage{float}
\usepackage{tablefootnote}
\usepackage{threeparttable}
\usepackage{ulem}

\author{Gonzalo Díaz Mirón}
\affiliation[ICTP]{Condensed Matter and Statistical Physics, The Abdus Salam International Centre for Theoretical Physics, 34151 Trieste, Italy}
\email{gdiaz_mi@ictp.it}
\author{Carlos R. Lien-Medrano}
\affiliation[Bremen]{Institute for Theoretical Physics and Bremen Center for Computational Materials Science, University of Bremen, 28359 Bremen, Germany}
\author{Debarshi Banerjee}
\affiliation[ICTP]{Condensed Matter and Statistical Physics, The Abdus Salam International Centre for Theoretical Physics, 34151 Trieste, Italy}
\alsoaffiliation[SISSA]{Scuola Internazionale Superiore di Studi Avanzati (SISSA), 34136 Trieste, Italy}
\author{Marta Monti}
\affiliation[ICTP]{Condensed Matter and Statistical Physics, The Abdus Salam International Centre for Theoretical Physics, 34151 Trieste, Italy}
\author{B. Aradi}
\affiliation[Bremen]{Institute for Theoretical Physics and Bremen Center for Computational Materials Science, University of Bremen, 28359 Bremen, Germany}
\author{Michael A. Sentef}
\affiliation[Bremen]{Institute for Theoretical Physics and Bremen Center for Computational Materials Science, University of Bremen, 28359 Bremen, Germany}
\alsoaffiliation[MPSD]{Max Planck Institute for the Structure and Dynamics of Matter, Center for Free-Electron Laser Science (CFEL), 22761 Hamburg, Germany}
\author{Thomas A. Niehaus}
\affiliation[LYON]{Univ Lyon, Université Claude Bernard Lyon 1, CNRS, Institut Lumière Matière, F-69622 Villeurbanne, France}
\author{Ali Hassanali}
\affiliation[ICTP]{Condensed Matter and Statistical Physics, The Abdus Salam International Centre for Theoretical Physics, 34151 Trieste, Italy}
\email{ahassana@ictp.it}

\title[]{Trajectory Surface Hopping with Tight Binding Density Functional Theory applied to Molecular Motors}

\abbreviations{TSH, DFTB, Molecular motors}

\SectionNumbersOn

\begin{document}

\newpage
\begin{abstract}
Non-adiabatic molecular dynamics (NAMD) has become an essential computational technique for studying the photophysical relaxation of molecular systems after light absorption. These phenomena require approximations that go beyond the Born-Oppenheimer approximation, and the accuracy of the results heavily depends on the electronic structure theory employed. Sophisticated electronic methods, however, make these techniques computationally expensive, even for medium size systems. Consequently, simulations are often performed on simplified models to interpret experimental results.

In this context, a variety of techniques have been developed to perform NAMD using approximate methods, particularly Density Functional Tight Binding (DFTB). Despite the use of these techniques on large systems where ab initio methods are computationally prohibitive, a comprehensive validation has been lacking. In this work, we present a new implementation of trajectory surface hopping (TSH) combined with DFTB, utilizing non-adiabatic coupling vectors (NACVs). We selected two different systems for validation, providing an exhaustive comparison with higher-level electronic structure methods.

As a case study, we simulated a system from the class of molecular motors, which has been extensively studied experimentally but remains challenging to simulate with ab initio methods due to its inherent complexity. Our approach effectively captures the key photophysical mechanism of dihedral rotation after absorption of light. Additionally, we successfully reproduce the transition from the bright to dark states observed in the time dependent fluorescence experiments, providing valuable insights into this critical part of the photophysical behavior in molecular motors.

\end{abstract}


\newpage
\section{Introduction}

In recent years, the development of non-adiabatic molecular dynamics (NAMD) simulations have greatly enhanced our understanding of complex photophysical and photochemical processes\cite{palombo2022retinal,jones2022steric,miron2023carbonyl,fregoni2018manipulating,filatov2022towards,mai2016origin}. These simulations are crucial for studying phenomena where electronic and nuclear degrees of freedom are strongly coupled, such as radiationless transitions and photoreactivity\cite{crespo2018recent,curchod2018ab,barbatti2011nonadiabatic}. However, traditional NAMD approaches can be computationally prohibitive due to the cost associated with advanced electronic structure methods and the need for extensive simulations to achieve well-converged results. Consequently, innovative strategies have emerged to integrate approximate electronic methods with NAMD \cite{pieroni2024investigating,granucci2001direct,kazaryan2010understanding}.

Among these strategies, several implementations have combined NAMD with Density Functional Tight-Binding (DFTB)\cite{akimov2016nonadiabatic,pal2016nonadiabatic,shakiba2022nonadiabatic,humeniuk2017dftbaby,stojanovic2017nonadiabatic,Wu2022NEXMDDFTB}, each exhibiting varying degrees of success and inherent limitations. These implementations typically avoid the analytic calculation of the Non-Adiabatic Coupling Vectors (NACVs) by employing either a Landau-Zener approach\cite{díaz2024exploring} or by calculating Time-Dependent Non Adiabatic Coupling (TD-NAC) using numerical methods\cite{hammes1994proton,ryabinkin2015fast}.
The TD-NAC approach has certain limitations, as it requires small time steps (less than 0.5 fs)\cite{wang2016recent,tran2019mechanisms} to prevent numerical issues.
Additionally, its computation depends on the overlap matrix between conformations at two consecutive time steps.
In DFTB, this matrix is parameterized for pairs of atoms at specific inter-atomic distances which are often greater than those encountered in NAMD simulations. 
All aforementioned limitations can be overcomed through the analytical calculation of the NACVs\cite{barbatti2011nonadiabatic,crespo2018recent}. 

In light of these challenges, we present an implementation of non-adiabatic dynamics employing the DFTB method\cite{elstner1998self}, utilizing the open-source codes DFTB+\cite{hourahine2020dftb+} and SHARC\cite{mai2018nonadiabatic,plasser10sharc3}. Our approach integrates the analytical NACVs developed by Send et al.\cite{send2010first} and implemented in Time Dependent DFTB (TD-DFTB) by Niehaus \cite{niehaus2021ground,niehaus2023exact}. This method provides a more accurate and stable description of electronic transitions compared to traditional TD-NAC approaches, particularly for studying photophysical mechanisms involving conical intersections (CoIns) between $S_1$ and $S_0$ states\cite{send2010first}. 

The primary goal of this work is to perform a detailed validation of our approach against previous studies using high-level electronic structure calculations. Previous implementations of NAMD within the DFTB method\cite{akimov2016nonadiabatic,pal2016nonadiabatic,shakiba2022nonadiabatic,stojanovic2017nonadiabatic} have focused directly on large systems, where \textit{ab initio} methods are unfeasible. This leads to a gap in validation against known benchmarks. To address this, we used two systems with very well-known photophysical mechanisms, namely the methaniminium cation and furan. Although these systems are small, their complex photophysics make them excellent benchmarks for validating our approach\cite{barbatti2006ultrafast,filatov2021signatures}.

We then apply our techniques to the photophysics of molecular motors\cite{garcia2019light,roy2024excited}. This is a very active field, which was awarded the Nobel Prize in Chemistry in 2016. Molecular motors can perform mechanical work using light as a stimulus, with applications ranging from nanotechnology to biology\cite{coskun2012great,omabegho2009bipedal,garcia2019light}. It is known from experiments combined with theory that upon photo-excitation, there is an ultrafast rotation of molecules along the dihedral angle, which is crucial for their operation as motors. Previous computational studies have used \textit{ab initio} methods for small molecular motors\cite{filatov2022towards}, a limited set of NAMD simulations\cite{wen2023excited}, or semi-empirical methods\cite{kazaryan2010understanding,kazaryan2011surface,pang2017watching}. Although the broad outlines of the mechanism are well-established, much work remains to be done, especially in understanding the specific details\cite{roy2024excited} such as the influence on the photophysics of substituents on the system\cite{conyard2014chemically}, polarity and viscosity of the solvent\cite{conyard2014chemically}, as well as the generation of charge transfer states\cite{roy2023control}, aspects that experiments alone cannot provide at an atomistic level. Our work not only further validates our method but also provides insights into the photophysical mechanisms . In particular, the low computational cost associated with our approach allow us to quantify how the vibrational modes modulate the different properties in the excited states of molecular motors.


\newpage
\section{Theoretical Approaches}

In this section, we summarize the key aspects of the DFTB theory and its time-dependent extension to the frequency domain, TD-DFTB. We also describe important aspects of non-adiabatic dynamics in the context of trajectory surface hopping (TSH). For a more detailed description of theory and implementation of DFTB and TSH algorithms, readers will be directed to relevant bibliographic references throughout the text.

\subsection{Density Functional Tight-Binding (DFTB and TD-DFTB)}

DFTB equations are derived from DFT by expansion of the total energy in a Taylor series of the electron density fluctuations $\delta\rho$ around a reference density $\rho^0$\cite{hourahine2020dftb+}. The chosen reference density is usually a summation of overlapping, spherical, and non-interacting atomic charge densities. Up to the second order (DFTB2), the total energy can be approximated as:\cite{Gaus2011}

\begin{equation}
\begin{split}
    E^{\text{DFT}}[\rho^0+\delta\rho] &\approx E^0[\rho^0] + E^{1\text{st}}[\rho^0,\delta\rho] + E^{2\text{nd}}[\rho^0,(\delta\rho)^2] \\
    E^{\text{DFTB2}}[\rho^0+\delta\rho] &= \sum_{A>B}E_{AB}^{\text{rep}} + \sum^{\text{occ.}}_i n_i \langle{\psi_i}|\mathcal{H}[\rho^0]|\psi_i\rangle + \frac{1}{2} \sum_{AB} \gamma_{AB} \Delta q_A \Delta q_B
    \label{eq:tbdft}
\end{split}
\end{equation}
where, $E^{rep}_{AB}$ is a pairwise repulsive potential energy, $|\psi_i\rangle$ is the $i^{\text{th}}$ molecular orbital in the linear combination of atomic orbitals (LCAO) framework, $n_i$ is the occupation, $\mathcal{H}[\rho^0]$ is the Hamiltonian operator in a two-center approximation, $\Delta q_A$ is the Mulliken charge on atom $A$ and $\gamma_{AB}$ represents the electron interaction of two Slater-type spherical charge densities on atoms $A$ and $B$. The inclusion of the second-order term requires a self-consistent solution since the Mulliken charges depend on the molecular orbitals\cite{hourahine2020dftb+}. 

For the calculation of the properties of the excited states, we employed the Casida formalism\cite{casida2009time} within the framework of DFTB\cite{niehaus2001tight,niehaus2009approximate}. The electronic excitation energy $\Omega_{I}$ in TD-DFTB can be obtained by solving the following eigenvalue problem:

\begin{equation}
 \begin{pmatrix}
   \textbf{A} & \textbf{B}\\
   \textbf{B} & \textbf{A}
 \end{pmatrix}
 \begin{pmatrix}
   \textbf{X} \\
   \textbf{Y} 
 \end{pmatrix} 
 = \Omega
 \begin{pmatrix}
   \textbf{1} & \textbf{0}\\
   \textbf{0} & \textbf{-1}
 \end{pmatrix}
 \begin{pmatrix}
   \textbf{X} \\
   \textbf{Y} 
 \end{pmatrix} 
  \label{eq:casida}
\end{equation}
where \textbf{X}, \textbf{Y} determine the transition density and oscillator strength, $\Omega$ denotes the transition energy of an associated excited state and the matrices \textbf{A} and \textbf{B} take the following form:
\begin{equation}
\begin{split}
  A_{ia\sigma,jb\tau} &= \dfrac{\delta_{ij}\delta_{ab}\delta_{\sigma\tau}(\epsilon_a-\epsilon_i)}{\eta_{j\tau}-\eta_{b\tau}} + K_{ia\sigma,jb\tau} \\
  B_{ia\sigma,jb\tau} &= K_{ia\sigma,bj\tau}
  \label{eq:matrix}
\end{split}
\end{equation}
where $\epsilon_{i}$ is the orbital energy, $i,j$ and $a,b$ are occupied and unoccupied KS orbitals respectively, $\sigma$ and $\tau$ are spin indices, whereas $K$ is the so-called coupling matrix and under monopole approximation can adopt simple expressions thereby reducing the computational cost\cite{niehaus2009approximate}. 

Gradients for both ground and excited states as well as oscillator strengths in the context of DFTB follow the same procedure used for DFT/TD-DFT\cite{sokolov2021analytical}. The most important feature in the present work are the calculation of analytical non-adiabatic coupling vectors.

\begin{equation}
\begin{split}
    \Omega_{nm}\mathbf{d_{nm}} &= \sum_{\mu\nu\sigma} \left(\dfrac{\partial H^0_{\mu\nu}}{\partial \xi} P_{\mu\nu\sigma}^{nm} - \dfrac{\partial S_{\mu\nu}}{\partial \xi} W_{\mu\nu\sigma}^{nm}\right) + \sum_{\mu\nu\sigma,\kappa\lambda\tau} \dfrac{\partial (\mu\nu|\nu_C + f^{xc,\omega}_{\sigma\tau}|\kappa\lambda)}{\partial \xi} \Gamma_{\mu\nu\sigma,\kappa\lambda\tau}^{nm} \\
    & + \sum_{\mu\nu\sigma,\kappa\lambda\tau} \dfrac{\partial (\mu\nu|\nu_C^{lr,\omega}|\kappa\lambda)}{\partial \xi} \Gamma_{\mu\nu\sigma,\kappa\lambda\tau}^{lr,nm} 
    \label{eq:tdnacv}
\end{split}
\end{equation}
where $n,m$ denotes different electronic states, $\Omega$, $H^0$ and $S$ are the excitation energy, the zero-order hamiltonian operator and the overlap matrix, respectively. $P$, $W$ and $\Gamma$ are the relaxed one-particle difference, the energy-weighted, and the two-particle density matrices, respectively. The last two terms refer to the functional and the long-range components. Expression for the different matrices in Eq \ref{eq:tdnacv} can be found in Ref.\citenum{niehaus2023exact}.

In this work, we also used the Tamm-Dancoff Approximation (TDA)\cite{hirata1999time}, which involves setting \textbf{B} and \textbf{Y} to zero in Equations \ref{eq:casida}-\ref{eq:tdnacv}. 
It is well known that TDA give similar results to Casida and alleviates the instability problem near the conical intersections\cite{hu2014performance,li2014configuration}.
To this end, we implemented the TDA as a new feature in the DFTB+, which will be available in an upcoming release of the code.

\subsection{Trajectory Surface Hopping (TSH)}

In each TSH trajectory, the electronic states are propagated using quantum mechanics, while the nuclear motion is handled classically, employing the forces coming from a single potential energy surface from the electronic structure method. At each timestep, the non-radiative probability is computed and a stochastic algorithm is employed to decide which potential energy surface the system will proceed along\cite{nelson2020non,curchod2018ab,barbatti2011nonadiabatic}. This probability is expressed as:

\begin{equation}
    P_{nm}(t) = -2 \int_t^{t+\delta t} dt' \dfrac{C_n(t')C_m^*(t')}{C_n(t')C_m^*(t')} \left(\mathbf{V}(R,t') \cdot \mathbf{d_{nm}}(R,t')\right)
    \label{eq:tully}
\end{equation}
where $C_n$ are the electronic coefficients of the potential energy surface $n$, $\mathbf{V}$ is the nuclear velocity and $\mathbf{d_{nm}}$ are the non adiabatic coupling vectors (NACVs) between the states $n,m$ (Eq \ref{eq:tdnacv}).

Since the photophysical relaxation of the system occurs on a very short timescale (from femtoseconds to picoseconds), the trajectory is highly correlated with the initial conditions. Therefore, an ensemble of trajectories should be run to ensure good convergence of the results. Additionally, a decoherence correction using the method developed by Granucci et al. was added\cite{granucci2007critical}.

\section{Computational Details}

\begin{figure}[H]
    \centering
    \includegraphics[width=0.97\linewidth]{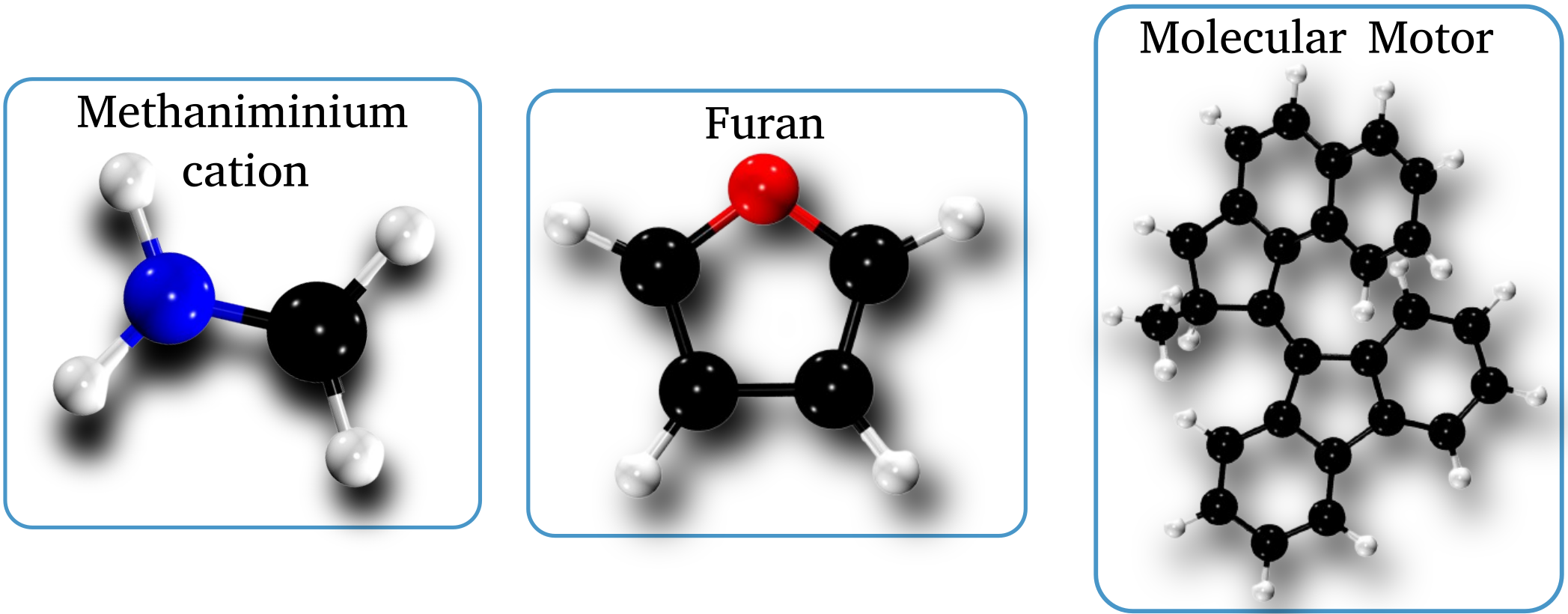}
    \caption{Systems studied in this work. Molecular motor is a generic name for the compound 9-(2,4,7-trimethyl-2,3-dihydro-1H-inden-1-ylidene)-9H-fluorene and it will be used throughout the text. Carbon, Oxygen, Nitrogen, and Hydrogen atoms are specified in black, red, blue and white colors, respectively.}
    \label{fig:allSYS}
\end{figure}

In this section, we describe the protocol employed to conduct non-adiabatic simulations for all the systems investigated in the present work shown in Figure \ref{fig:allSYS}. For all simulations reported herein, the DFTB+ (version 24.1) code was employed for the quantum mechanical calculations, while the SHARC code (version 3.0) performed the non-adiabatic dynamics and facilitated various stages of preparatory and subsequent analysis. To benchmark the results obtained with DFTB and/or TD-DFTB against higher-level electronic structure methods, we utilized the ORCA electronic structure code\cite{neese2012orca}.

The simulation protocol begins with the geometry optimization of the molecular system in vacuum. Upon completion of the optimization, a frequency calculation is performed to obtain the normal modes and vibrational frequencies of the system. Using these, 200 random initial conditions (positions and velocities) are generated by sampling from the Wigner distribution at a temperature of 300K. For all the initial conditions, a TD-DFTB calculation was performed including 10 excitations. The final absorption spectra for each system were computed by averaging and applying a Gaussian function with a full width at half maximum (FWHM) of 0.2 eV.

Non-adiabatic molecular dynamics (NAMD) simulations were conducted for all initial conditions in the NVE ensemble with a timestep of 0.5 fs. The hopping events were determined based on the non-adiabatic coupling vectors, as detailed in the Theoretical Approaches section. We applied a decoherence parameter of 0.1 hartree\cite{granucci2007critical}. The simulations were run for 200 fs including three electronic states for the methaniminium cation, 200 fs with two electronic states for furan and 1 ps with three electronic states for the molecular motor system.

The variations between the systems involve the setup of the quantum calculations inside DFTB+ code. For the methaniminium cation and furan systems, we utilized DFTB up to second-order corrections (DFTB2) and TD-DFTB with the Casida algorithm, employing the mio Slater-Koster parameters\cite{gaus2011dftb3}. For the molecular motor system, we employed DFTB2 and TD-DFTB with the Tamm-Dancoff Approximation (TDA) and Long-Range Corrections (LC)\cite{lutsker2015implementation,kranz2017time,sokolov2021analytical} using the ob2 Slater-Koster parameters\cite{vuong2018parametrization}.

For the molecular motor system we calculated the time-dependent fluorescence emission using data from all NAMD simulations, up until the point where the system undergoes non-radiative decay to the ground state, it means when the system is in the excited state only. At each time step, the fluorescence spectra were generated using the oscillator strength and the energy gap, and averaged across the different trajectories. Additionally, we applied a 50 fs time-averaging window to the spectra to simulate the experimental time resolution.

\section{Results and Discussion}

\subsection{Methaniminium cation}

This system has been extensively studied using a wide range of electronic structure methods\cite{hollas2018nonadiabatic,suchan2020pragmatic,bonavcic1987critically,barbatti2006ultrafast}. Despite its simplicity and small size, the molecule exhibits notably complex photophysics, making it an interesting system to test our non-adiabatic molecular dynamics (NAMD) simulations based on TD-DFTB. The process involves a three-state problem, including the ground state and two electronic excited states: $\pi\sigma^* (S_1)$ and $\pi\pi^* (S_2)$. 

\begin{figure}[H]
    \centering
    \includegraphics[width=1.0\linewidth]{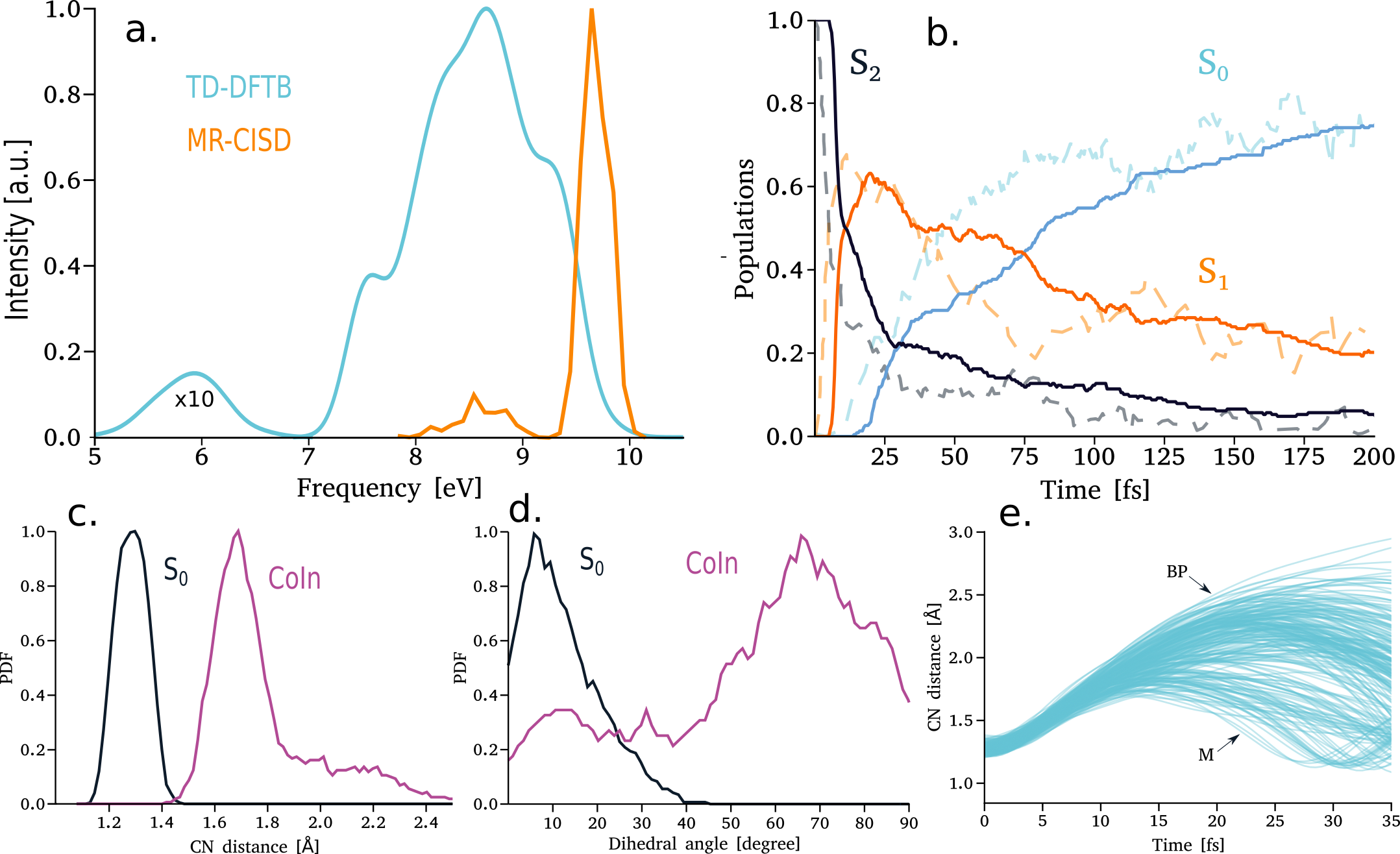}
    \caption{Photophysical mechanism of methaniminium cation. \textbf{Panel a:} Average absorption spectra calculated from all the initial conditions employed for NAMD using TD-DFTB (blue line) and using MR-CISD\cite{barbatti2007fly} (orange line). The first band was increased by a factor of x10 for visualization. \textbf{Panel b:} Temporal evolution obtained from all the NAMD simulations. Results for TD-DFTB are shown in solid lines and for MR-CISD\cite{barbatti2007fly} are shown in dashed lines.\textbf{Panel c:} DFTB Histograms of the CN distances at geometries in $S_0$ in the initial conditions and at the CoIn $S_2/S_1$. \textbf{Panel d:} DFTB Histograms of the dihedral angle at geometries in $S_0$ in the initial conditions and at the CoIn $S_1/S_0$. \textbf{Panel e:} Temporal evolution of the CN distance for all the NAMD simulations obtained with DFTB. The bipiramidalization (BP) and mixed (M) branches are marked in the plot. All the results for MR-CISD, which stands for MR-CISD/SA-3CAS(4,3)/6-31G*, were extracted from references \citenum{barbatti2007fly,barbatti2006ultrafast}.}
    \label{fig:formal}
\end{figure}

Figure \ref{fig:formal} presents the results generated using our DFTB approach and compares them with the high-level electronic method MR-CISD\cite{barbatti2007fly,barbatti2006ultrafast}. Panel a shows that the TD-DFTB absorption spectra predicts lower energies for the excited states and a higher energy gap $S_2\to S_1$ compared to a more accurate method (see Table S1 in the supplementary information).

Despite these differences, panel b in Figure \ref{fig:formal} shows that the average electronic populations over time from TD-DFTB simulations align well with those from MR-CISD\cite{barbatti2007fly}, although the predicted lifetimes are slightly longer, attributed to the higher energy gap between $S_2\to S_1$ predicted by TD-DFTB. Table S2 in the supplementary information compares the lifetimes for both electronic states obtained using our approach with those derived from various sophisticated methods. 

Overall, TD-DFTB provides good qualitative insights into the photorelaxation mechanism, in excellent agreement with other methods. It is well known that the conical intersection (CoIn) between $S_2/S_1$ involves an stretching of the CN distance, while the CoIn $S_1/S_0$ proceeds through dihedral rotation\cite{barbatti2006ultrafast}. This is visualized in Figure \ref{fig:formal}, where we show the distribution for both modes on the ground state and at the respective CoIn (panels c and d). Figure S1 in the supplementary information shows the superposition of the CoIns optimizations using our method and its comparison with MR-CISD\cite{barbatti2006ultrafast}. Moreover, our approach successfully identify the branches, BP (BiPyramidalization) and M (Mixed) at the beginning of the photophysical decay\cite{barbatti2007fly} (panel e). 

The success of our approach in predicting the most important features of the photophysical relaxation is due to the ability of TD-DFTB to capture two critical factors: the ordering of the electronic states and their electronic properties in terms of the molecular orbitals compositions, even if the energy values are shifted compared to more sophisticated electronic methods.

\subsection{Furan}

Furan is known to have two lower-lying excited states at the Franck-Condon (FC) region: a Rydberg state, $\pi 3s$, and a $\pi\pi^*$ state\cite{filatov2021signatures,fuji2010ultrafast,oesterling2017substituent}. Within the DFTB method, Rydberg states are not predicted due to the use of a minimal basis set\cite{niehaus2021ground}. However, the Rydberg state does not significantly influence the photophysics of furan, allowing us to use DFTB to effectively study its mechanisms, as we will demonstrate below.

\begin{figure}[H]
    \centering
    \includegraphics[width=1.\linewidth]{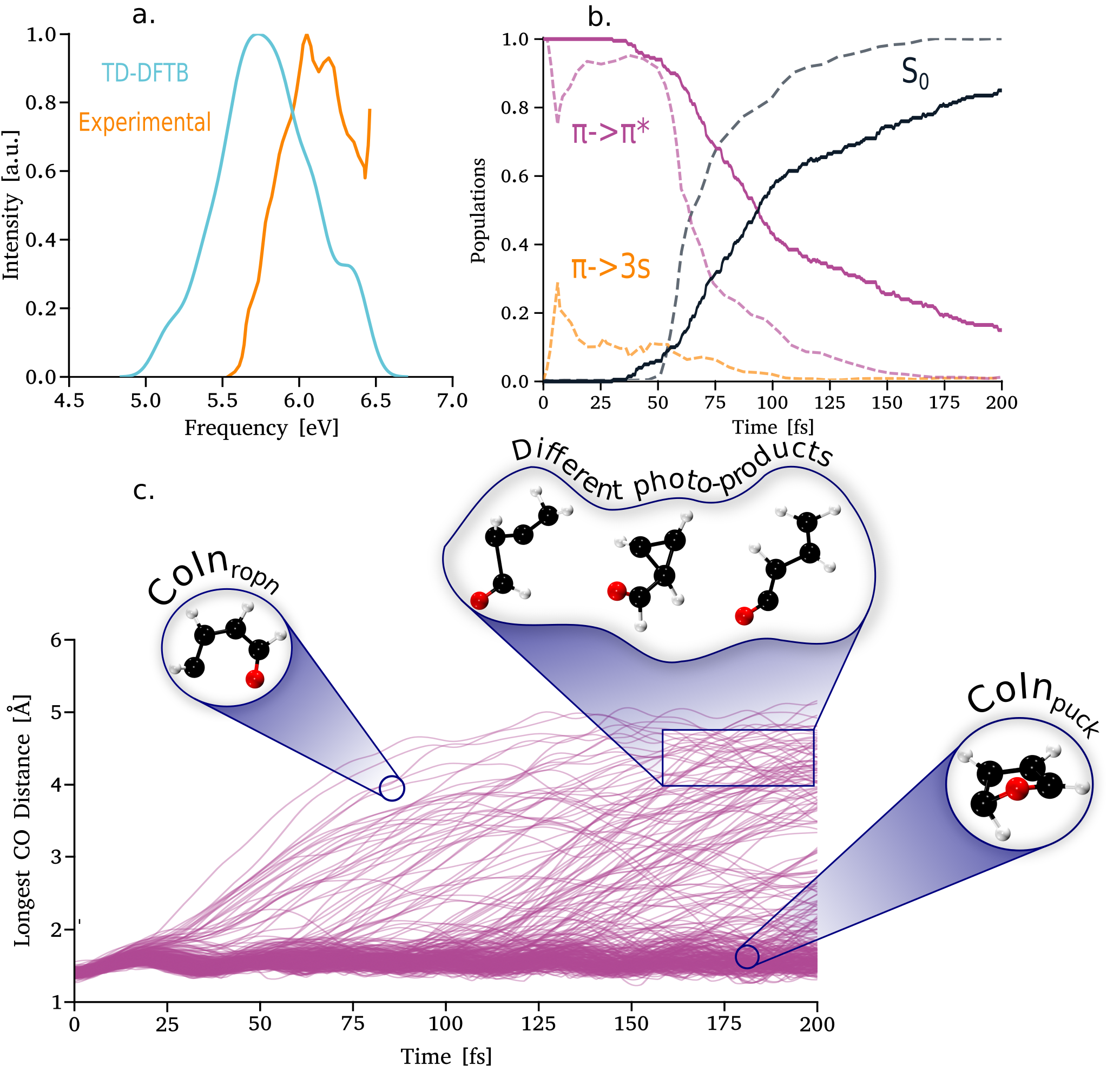}
    \caption{Photphysical mechanism of Furan. \textbf{Panel a:} Average absorption spectra calculated for all NAMD initial conditions using TD-DFTB (blue line) and its comparison with the experimental data (orange line) extracted from reference\cite{fuji2010ultrafast}. \textbf{Panel b:} Temporal evolution of the average populations of the excited states using all the NAMD simulations with our techniques (solid lines) and its comparison with the adiabatic populations calculated with TD-DFT/PBE0 (dashed lines), extracted from reference\cite{fuji2010ultrafast}. \textbf{Panel c:} Temporal evolution of the longest CO distance for all the NAMD simulations based on DFTB (purple line). The two different conical intersections of furan are highlighted in blue circles (CoIn$_{ropn}$ and CoIn$_{puck}$). We also highlight the different photo-products obtained \textit{via} CoIn$_{ropn}$.}
    \label{fig:furan}
\end{figure}

A previous study performing static calculations on the furan molecule showed that TD-DFTB aligns well with higher-level electronic methods\cite{niehaus2021ground}. However, understanding the photophysical relaxation of the system requires the incorporation of dynamical effects, which will be the focus of the present work. Panel a in Figure \ref{fig:furan} shows the absorption spectra and its comparison with the experimental data. Our results show good agreement with the experiment, with an error less than 0.5 eV. This indicates that the main contribution to the absorption spectrum comes from the $\pi\pi^*$ state\cite{fuji2010ultrafast} (see Figure S2 in the supplementary information).

Panel b of Figure \ref{fig:furan} shows the evolution of the population for the electronic states using our approach in its comparison with TD-DFT/PBE0-6-311++G**\cite{fuji2010ultrafast}. The lifetime predicted by our approach is approximately 20 fs longer than the one observed in TD-DFT. However, quantum dynamics methods\cite{gromov2004theoretical,gromov2011ab} have also predicted longer lifetimes, demonstrating good agreement with our approach.

Our approach successfully predicts the two well-known conical intersections\cite{filatov2021signatures, oesterling2017substituent}: CoIn$_{puck}$ (puckering, an out-of-plane vibration) and CoIn$_{ropn}$ (ring-opening). Panel c in Figure \ref{fig:furan} shows the longest CO bond distance across all the NAMD simulations using our approach, where we highlighted both conical intersections and the variety of photo-products obtained.

A more quantitative comparison of lifetimes and the percentage of the system reaching each of the CoIn using our approach and with higher-electronic methods can be found in Table S3 in the supplementary material. These results demonstrate that our TD-DFTB/NAMD simulations align well with other apporaches using more sophisticated electronic theory levels\cite{fuji2010ultrafast,filatov2021signatures,oesterling2017substituent}.

\subsection{Molecular Motors}

In the previous sections, we demonstrated that our method based in DFTB is effective for studying complex photophysical phenomena. With this validation, we next move to investigating the molecule 9-(2,4,7-trimethyl-2,3-dihydro-1H-inden-1-ylidene)-9H-fluorene, which belongs to the class of overcrowded molecular motors\cite{garcia2019light}. These systems are excellent candidates for using DFTB because their medium-to-large size makes the application of \textit{ab initio} methods computationally demanding. However, there are a few theoretical studies employing both \textit{ab initio} and semi-empirical methods\cite{pang2017watching,wen2023excited} on this system, which provide useful results for validating our approach.

\begin{figure}[H]
    \centering
    \includegraphics[width=1.\linewidth]{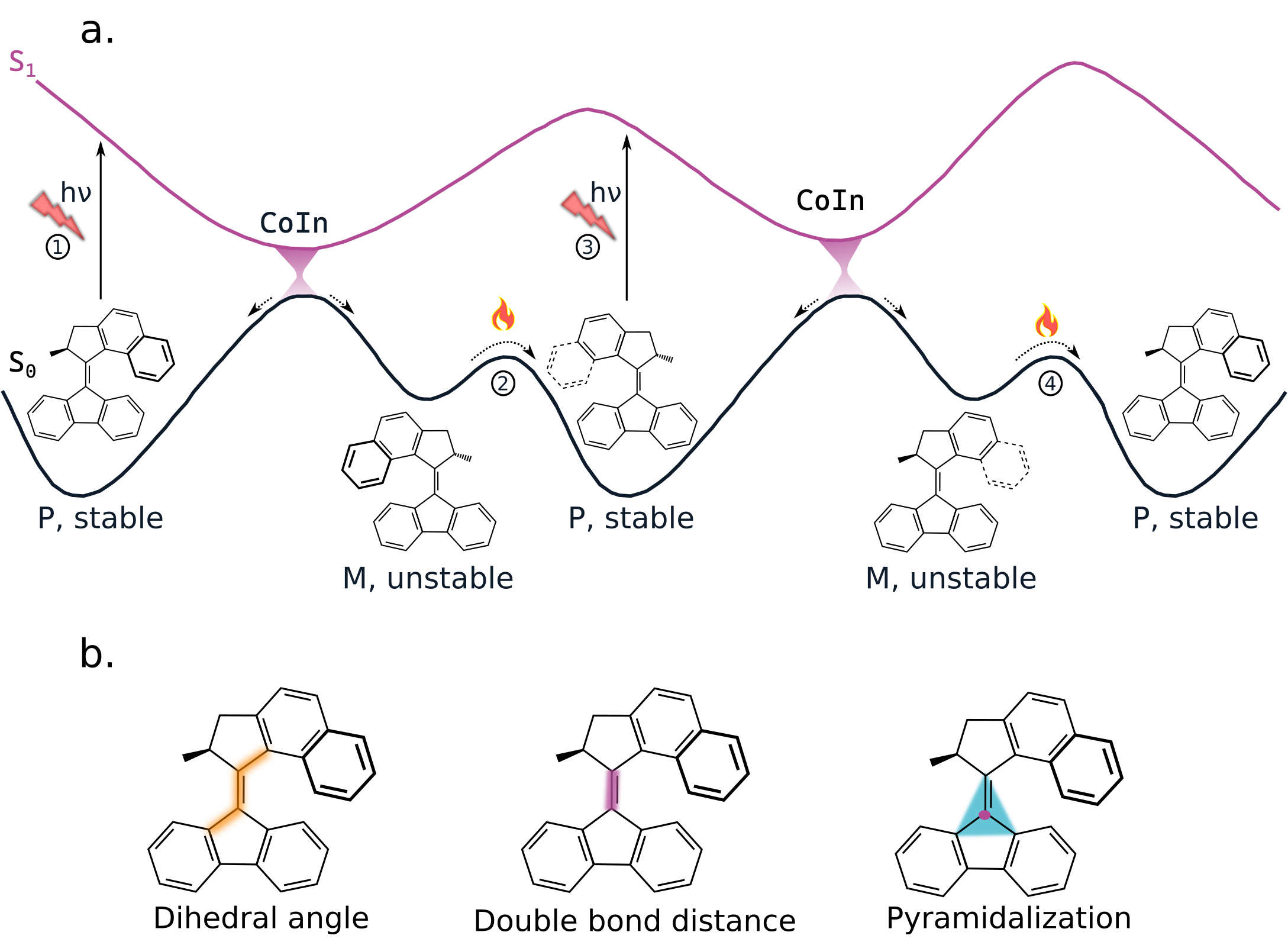}
    \caption{\textbf{Panel a:} Schematic representation of the four cycle steps in the molecular motors mechanism. \textbf{Panel b:} Molecular motor geometry highlighting the different modes used to describe the photophysical mechanism.}
    \label{fig:mecMot}
\end{figure}

To provide some context, we will first briefly describe the photochemistry of these systems. More detailed information can be found in excellent reviews on this topic\cite{garcia2019light,roy2024excited}. Panel a in Figure \ref{fig:mecMot} shows the four steps during a cycle of the molecular motor. The first part of the mechanism involves a photon-energy step (1), where the molecular motor absorbs light, causing a change in its electronic state. This absorption induces a rotation in the dihedral angle, resulting in the formation of an unstable conformer (M) in the ground electronic state through the passage of the conical intersection (CoIn). Following this, thermal energy facilitates a helix inversion (2), leading to the generation of a stable conformer (P). The initial photon-energy step is crucial because it provides the energy required to drive the motor, making it the primary focus of this study. 

Additionally, panel b in Figure \ref{fig:mecMot} highlights the principal modes involved in the photophysical mechanism in molecular motors and these will be used throughout the text. 

\begin{figure}[H]
    \centering
    \includegraphics{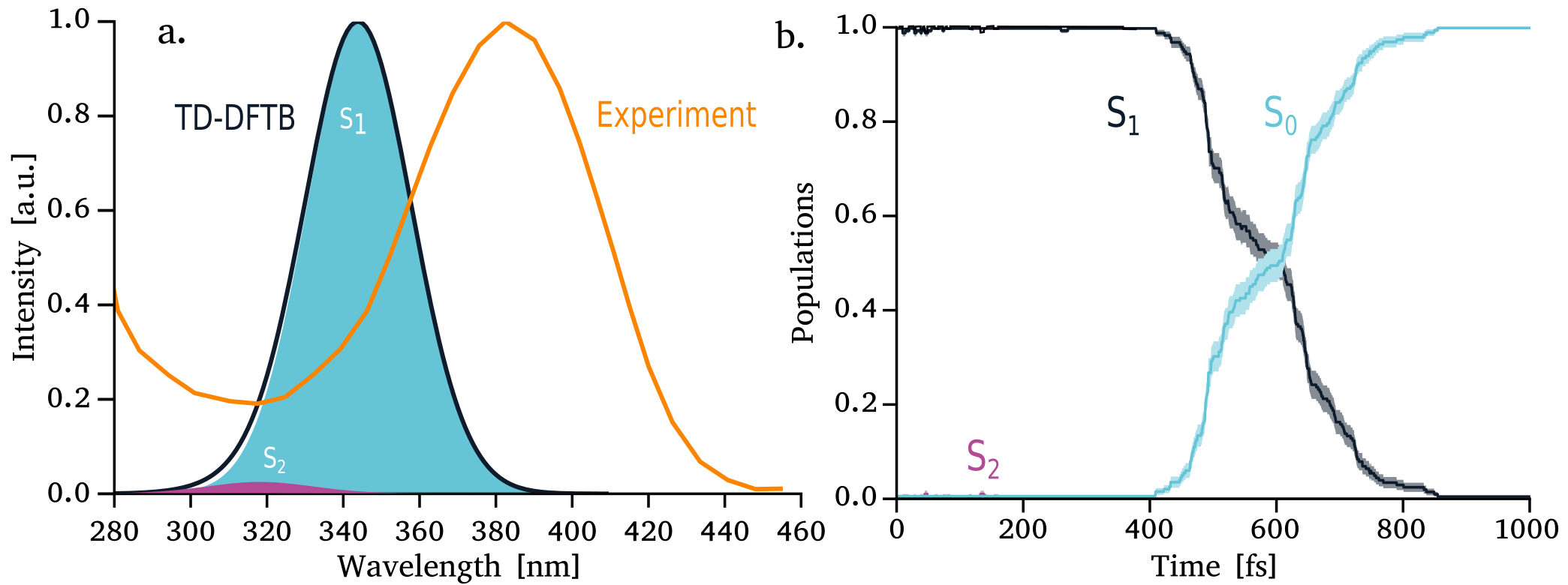}
    \caption{\textbf{Panel a:} Total average absorption spectra calculated using all the initial conditions for NAMD using TD-DFTB (black line), transitions to $S_1$ and $S_2$ are depicted as a blue and purple area, respectively. Experimental spectra, reproduced from reference\citenum{vicario2005controlling}, is shown in orange line. \textbf{Panel b:} Average populations of the electronic states as a function of time.}
    \label{fig:absMot}
\end{figure}

Before delving into the specific mechanism of non-radiative decay in the molecular motors, panel a in Figure \ref{fig:absMot} presents the average absorption spectra calculated using TD-DFTB for all the initial conditions. This is compared with the experimental results from the work of Vicario et. al.\cite{vicario2005controlling}

Panel a in Figure \ref{fig:absMot} shows that the primary contribution to the first absorption band arises from the excitation to the $S_1$ state, corresponding to a $\pi\to\pi^*$ transition around the central double bond. The molecular orbitals involved in the transition to $S_1$ are shown in Figure S3 in the supplementary information along with a comparison to those obtained using TD-DFT with LC (see Table S4 in the supplementary information for a comparison of TD-DFT without LC). These results align well with previous theoretical studies\cite{wen2023excited,pang2017watching}. However, the most important difference in this regard is that our approach predicts a blue-shift of 45 nm (0.4 eV) compared to the experimental maximum absorption. One possible explanation for this difference could be the absence of solvent in our simulations, while all the experiments discussed throughout the text are conducted in solution. 

\begin{figure}[H]
    \centering
    \includegraphics[width=1.0\linewidth]{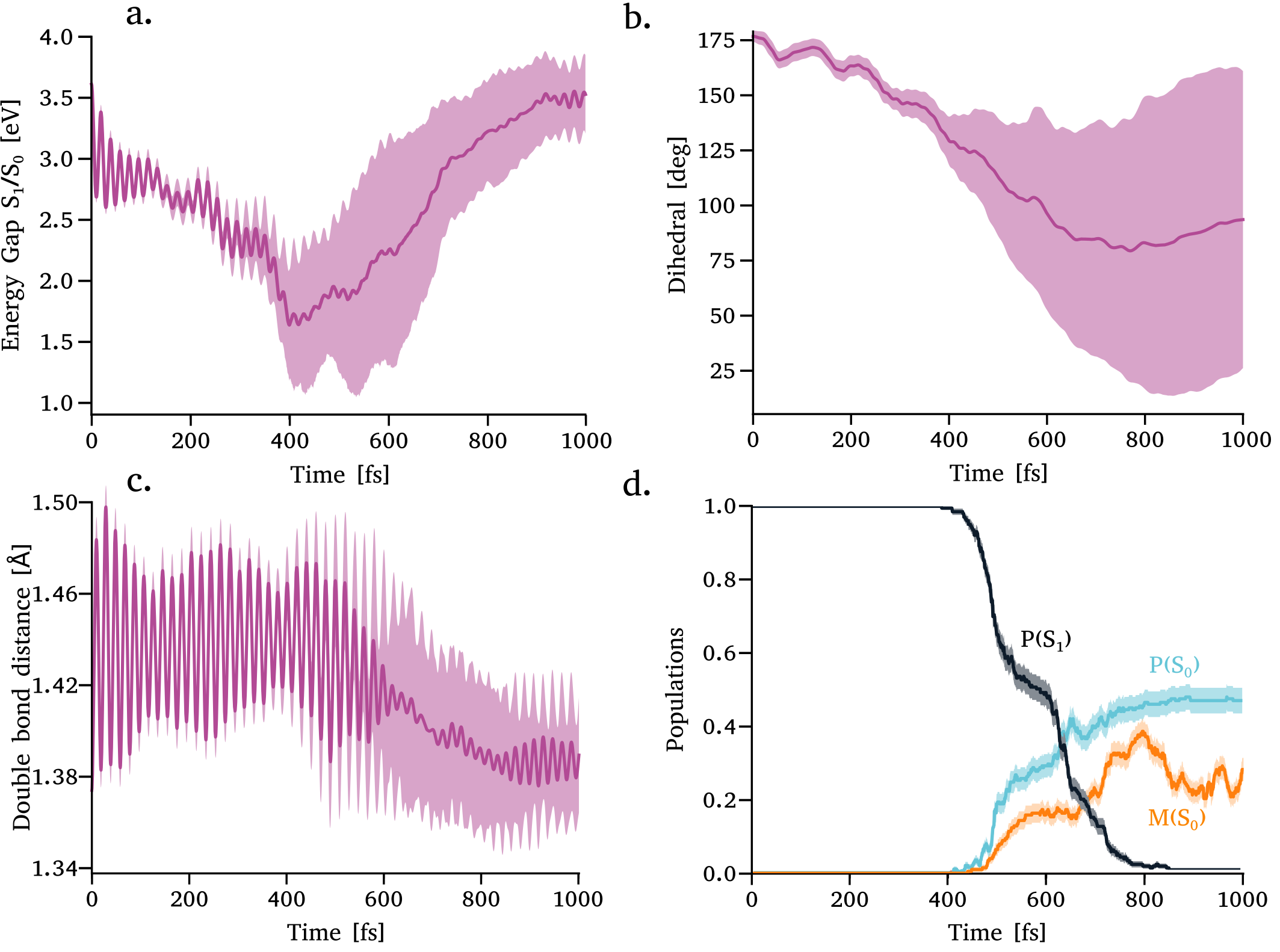}
    \caption{Non-radiative mechanism in molecular motors. \textbf{Panel a:} Energy gap $S_1\to S_0$ along time. \textbf{Panel b:} Dihedral angle along time. \textbf{Panel c:} Double bond distance along time. \textbf{Panel d:} Populations of the conformers along the time. All the averages values are presented in solid lines while the dispersion of the values as shadows area.}
    \label{fig:namdMot}
\end{figure}

Now, we turn into the photophysical mechanism. From panel b in Figure \ref{fig:absMot}, we observe a complete population $S_1\to S_0$ transfer before 1 ps, with an effective lifetime of 660 fs. This result aligns well with previous theoretical findings\cite{pal2016nonadiabatic,wen2023excited}. However, our approach differs from other studies regarding the waiting time (the time that takes to start observing the non-radiative decay). While they predicted a value of around 200 fs\cite{pang2017watching,wen2023excited,kazaryan2011surface}, we find a longer waiting time of approximately 400 fs. To better understand this difference, we analyzed the temporal evolution of the average energy gap between $S_1$ and $S_0$ (panel a), the dihedral angle (panel b), and the double bond distance (panel c) in Figure \ref{fig:namdMot}.

The analysis presented in Figure \ref{fig:namdMot} allows us to divide the waiting time into two distinct regimes: In the first regime, which dominates the initial 200 fs, we observe from the Pearson correlation that the double bond distance ($R=-0.63$) modulates more the oscillations of the energy gap than the others modes (see Figure S4 in the supplementary material). Along these first 200 fs, the energy gap shows only slight reduction and the dihedral angle only performs about three oscillations near the starting conformation, indicating that the system remains close to the Franck-Condon region. Previous theoretical works employing excited-state optimizations suggested that the pathway along the dihedral mode is barrierless\cite{kazaryan2011surface,pang2017watching}. 
However, in dynamic simulations, the evolution of the system can take a different route due to dynamical effects\cite{kazaryan2011surface}. Although our results seem to be aligned with previous studies, we cannot rule out the existence of a barrier along the decay pathway using our approach. Indeed, the steric effect between hydrogen atoms of the two parts of the motor when the dihedral angle starts to decrease is not well captured by DFTB due to the use of a minimal basis set\cite{elstner2006scc}. Similar problems were found in the cis-trans isomerizations of azo-compounds, where the authors demonstrated that including dispersion corrections or increasing the size of substituents could minimize this error\cite{poidevin2023evaluation,brandenburg2014accurate}.

In the second regime of the waiting time, observed during the 200–400 fs interval, we see a better correlation between the dihedral angle and the energy gap ($R=0.86$, see Figure S4 in the supplementary information), with a small contribution from the double bond distance ($R=-0.28$). During this time, the energy gap decreases with the dihedral angle demonstrating that this mode is the most relevant in the photophysical decay. However, the energy gap that needs to be reduced is significantly larger in our approach compared to experimental observations (see Figure \ref{fig:absMot}), leading to a longer waiting time.

Another notable feature is the significant dispersion observed in the dihedral angle, energy gap, and double bond distance (see panels a-c in Figure \ref{fig:namdMot}). This dispersion occurs after the system crosses the CoIn, as it can adopt two distinct conformers in the ground state\cite{sheng2023designing,conyard2012ultrafast}. To characterize these conformers, we classified the molecular geometries as follows: those with a dihedral angle $100\le \theta [deg]\le 180$ were identified as the \textbf{P} conformer, while those with $20\le \theta [deg] \le 70$ were classified as the \textbf{M} conformer (see Figure \ref{fig:mecMot}). Similar classifications have been used in the study of other compounds\cite{kazaryan2011surface}. Based on this, we constructed the plot in panel d of Figure \ref{fig:namdMot}, which tracks the formation of these conformers in their respective electronic states. We observe that the formation of both conformers occurs over slightly different timescales, a point that will be further discussed later. Our results indicate that the quantum yield for \textbf{P} $\to$ \textbf{M} isomerization fluctuates between 30\% and 40\% consistent with previous theoretical studies \cite{kazaryan2011surface} but slightly larger than the 14\% reported experimentally \cite{sheng2023designing}.

Previous experimental studies using time-dependent fluorescence and transient absorption spectroscopies have demonstrated that molecular motors exhibit a non-exponential signal decay, with an ultrafast and slow component in the order of femtoseconds and picoseconds, respectively\cite{conyard2012ultrafast,conyard2014chemically,filatov2022towards}. It has been shown that the emission intensity drops to zero (or nearly zero) at approximately 100 fs, which is significantly faster than the non-radiative decay from $S_1\to S_0$ through the CoIn. This observation suggests that the system is initially in a bright state at the Franck-Condon region but rapidly moves away, thus reaching conformations in $S_1$ where the emission intensity is very low. Pan et. al have performed a bi-exponential fitting on the population decay of $S_1$ obtained from NAMD simulations\cite{pang2017watching}. However, such analysis with the population only reveals two different timescales for the non-radiative transition $S_1\to S_0$, which does not correspond to the experimental results. The transition from the bright to dark state can not be captured by the populations obtained from the NAMD since no information about the intensity of the emission is included.

\begin{figure}[H]
    \centering
    \includegraphics[width=1.\linewidth]{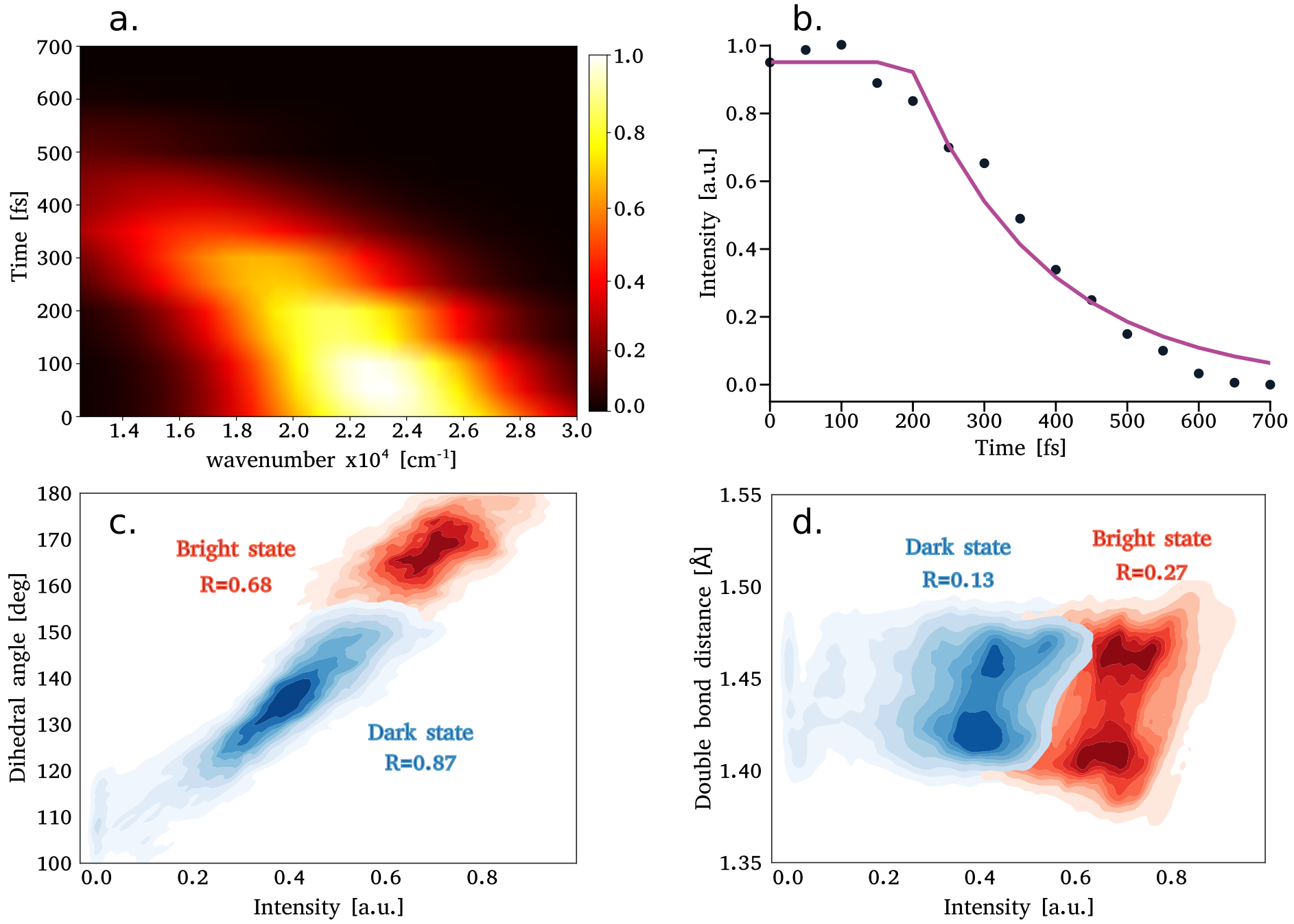}
    \caption{Time Dependent Fluorescence. \textbf{Panel a:} Time dependent fluorescence calculated from all the NAMD simulation including all the geometries that remain on $S_1$ state. The spectra was convoluted with a gaussian function with a FWHM of 1600 $cm^{-1}$ (approx 0.2 eV). The spectra was averaged every 50 fs, to simulate the experimental setup\cite{conyard2012ultrafast} \textbf{Panel b:} Temporal evolution of the intensity at the maximum emission (black dots), exponential fitting is shown in purple line. \textbf{Panel c:} 2D histogram of the distribution for the dihedral angle in both bright and dark states. \textbf{Panel d:} 2D histogram of the distribution for the double bond distance in both bright and dark states. Bright state was asssigned as the first 300 fs and the dark state between 350-600 fs of the simulation. The $R$ value represents the Pearson correlation factors between the modes and the emission intensity.}
    \label{fig:tdfluor}
\end{figure}

To demonstrate that our approach can capture this bright-to-dark state transition we calculated the average time-dependent fluorescence spectra using all the NAMD simulations in which the system remains in $S_1$. This result is presented in Figure \ref{fig:tdfluor}, panel a.

The calculated time dependent fluorescence spectrum in panel a of Figure \ref{fig:tdfluor} exhibits a pronounced red shift of approximately 7000 $cm^{-1}$, while the corresponding experimental value is 4000 $cm^{-1}$\cite{conyard2012ultrafast}. The discrepancy of approximately 3000 $cm^{-1}$ (0.37 eV) in our approach is comparable to that obtained for the absorption spectra (see Figure \ref{fig:absMot}). Additionally, we plotted the emission intensity as a function of time in panel b of Figure \ref{fig:tdfluor}. Fitting the intensity data reveals a lifetime of 340 fs. This lifetime is shorter than the one predicted using the population evolution for the radiationless transition $S_1\to S_0$ (660 fs, see panel b in Figure \ref{fig:absMot}), indicating that the system effectively reaches a dark state before crossing the CoIn. Remarkably, this observation is in agreement with experimental data\cite{conyard2012ultrafast}. However, as previously discussed, the time needed by the system to reach the dark state is longer in our DFTB approach compared to the experimental value (see below Figure \ref{fig:namdMot}).

There has been considerable debate regarding the interpretation of these results in terms of geometric conformations of the system. Experimental work by Conyard et al.\cite{conyard2012ultrafast} identified the pyramidalization mode as the primary factor responsible for generating this dark state. In contrast, theoretical studies\cite{wen2023excited,kazaryan2011surface} have suggested that this behavior is influenced by modes around the double bond, such as bond distance and dihedral angle. It is likely that all of these modes are coupled along the decay pathway, making it challenging to isolate individual contributions. Experiments are inherently limited in accessing atomistic details and theoretical studies require a large amount of data for a comprehensive analysis. Fortunately, the low computational cost associated with the DFTB approach allows us to generate extensive data on both bright and dark states, enabling a more detailed investigation into these complex interactions.

Panels c and d in Figure \ref{fig:tdfluor} display the 2D histograms for the dihedral angle and the double bond distance, respectively (pyramidalization mode is shown in Figure S5 in the supplementary information). The data is divided into two distinct phases: the bright state (red filled contour) occurring during the first 300 fs of the dynamics, and the dark state (blue filled contour), occurring between 350–600 fs. The criteria employed were based on the two distinct lifetimes. The bright state refers to the period before the system transitions to the dark state with a lifetime of 340 fs. The dark state is defined as the interval following this 340 fs and preceding the non-radiative $S_1\to S_0$ decay via the CoIn, which has a lifetime of 660 fs.

We can clearly see that the mode regulating the decrease in emission intensity is the dihedral angle. In the bright state, the Pearson correlation for the dihedral angle and bond distances are 0.68 and 0.27, respectively. In the dark state, the correlation for the dihedral mode increases to 0.87, while for the double bond distance, it decreases to 0.13. During the first few femtoseconds of the simulation, the system remains close to the Franck-Condon region for the dihedral mode while the double bond distance oscillates between 1.39 and 1.47 \AA. These changes in both modes do not produce a significant change in the oscillator strength (0.7-0.8 a.u.), confirming the presence of the bright state. However, after 350 fs, we observe that the oscillator strength decays from 0.6 a.u.\ to nearly zero, with a clear correlation to the dihedral mode. Comparable results were observed in the correlation analysis between the modes and the energy gap (see discussion below Figure \ref{fig:namdMot}). Similar conclusions can be drawn from the scatter density plot of both modes and emission intensity (see Figure S6 in the supplementary information).

Previous studies have suggested that the contribution to the fluorescence spectra when the system is in the dark state is due to coupling back to conformations of the system in the bright state\cite{conyard2014chemically}. Although there is a small overlap between the conformations for both states along the modes (panels c and d in Figure \ref{fig:tdfluor}), our results reveal that most of the contribution to the emission spectra comes from the dark state itself, but with considerably lower intensity.

With the previous analysis we can clearly see that our approach predicts the non-exponential decay, where we observed an ultrafast component for the bright $\to$ dark state with a lifetime of 340 fs and a slow decay for the transition $S_1\to S_0$ through the CoIn with a lifetime of 660 fs. The non-radiative transition through the CoIn occurs from the $S_1$ in the dark state. All of these results are in very good agreement with the experiments\cite{conyard2012ultrafast,conyard2014chemically}.Moreover, taking into account that the generation of the different conformers in the ground state occurs at slightly different time scales (panel d in Figure \ref{fig:namdMot}) we can conclude that there is a higher probability of generating the \textbf{M} conformer when the system takes more time to reach the $S_1\to S_0$ CoIn. This means that the system remains longer in the dark $S_1$ state. This result is consistent with the experimental study by Conyard et al.\cite{conyard2014chemically}, which found that by varying the substituents in the system, an increase in the lifetime of the dark state leads to an increase in the quantum yield of isomerization.

\section{Conclusions}

In this work, we have presented a new implementation for performing non-adiabatic molecular dynamics (NAMD) simulations using non-adiabatic coupling vectors (NACVs) within the framework of TD-DFTB theory. While there have been a few previous implementations using this theory, they have not fully utilized NACVs nor have they provided exhaustive validation against higher-level electronic structure methods. 

Our study demonstrates the effectiveness and accuracy of our DFTB approach across various systems by showing its ability to capture the most relevant features of photophysical relaxation, even when certain states do not perfectly match experimental data or higher-level methods. For example, in the methaniminium cation and furan system, DFTB accurately describes the complex photophysical deactivation, including the different conical intersections (CoIns) and the diverse photoproducts despite limitations such as not capturing the Rydberg state.

Our approach is particularly advantageous for studying molecular motors as it allows us to reproduce experimental values and results from other theoretical works with significantly lower computational cost. While the ADC(2) method require approximately 10.000 CPU hours for a single molecular dynamics simulation\cite{wen2023excited}, our method only requires 100 CPU hours, making it far more efficient. Compared to other semi-empirical methods, our approach does not suffer from the issue of tracking molecular orbitals which can be problematic in longer simulations\cite{kazaryan2011surface,pang2017watching}. This allows us to study photophysical relaxations over extended periods, providing deeper insights into the dynamics of molecular motors. With these advantages, we can study systems for longer times and larger sizes, such as DNA-based molecular motors\cite{omabegho2009bipedal,lubbe2018photoswitching}, which will be the subject of future applications of our approach.

We also provide valuable insights into the mechanisms of molecular motors. Due to the large number of NAMD simulations possible with our approach, we demonstrated how different modes modulate the energy gap between $S_1$ and $S_0$ and how they are coupled to the transition from the bright state to the dark state before reaching the conical intersection. Additionally, we verified experimental findings that showed that a long lifetime in the dark state results in an increased quantum yield of isomerization. These results are beneficial for studying how substituents and solvents affect photodynamics in these large systems.

In summary, our implementation of NAMD using TD-DFTB and NACVs provides a promising approach for studying complex photophysical phenomena. The ability to accurately simulate excited state molecular dynamics at a fraction of the computational cost of traditional methods opens new possibilities for exploring larger systems and longer timescales thereby expanding our understanding of molecular motors and other complex systems.

\begin{acknowledgement}

GDM, DB, MM and AH thank the European Commission for funding on the ERC Grant HyBOP 101043272. GDM also acknowledge CINECA supercomputing for the resource allocation (project NAFAA-HP10B4ZBB2). CRL-M acknowledge financial support from the German Research Foundation (DFG) through Grant No. FR 2833/82-1. MAS thank the ERC, Grant CAVMAT (project no. 101124492)

\end{acknowledgement}

\bibliography{bibref}

\end{document}


\begin{suppinfo}
\setcounter{table}{0}
\renewcommand{\thetable}{S\arabic{table}}
\newcounter{SItab}
\renewcommand{\theSItab}{S\arabic{SItab}}
\setcounter{figure}{0}
\renewcommand{\thefigure}{S\arabic{figure}}
\newcounter{SIfig}
\renewcommand{\theSIfig}{S\arabic{SIfig}}

\section{Methaniminium cation}

\begin{table}[H]
\begin{center}
  \begin{tabular}{ c c c c c c c }
    \hline
    States & This work & This work & Ref. A\cite{hollas2018nonadiabatic} & Ref. A\cite{hollas2018nonadiabatic}  & Ref. B\cite{barbatti2006ultrafast} & Ref. C\cite{bonavcic1987critically}\\
    \hline
    Method & (t) & (x) & (a) & (b) & (c) & (d) \\
    $S_0\to S_1 (\sigma\pi^*)$ & 6.19 & 7.71 & 9.17 & 9.27 & 8.35 & 8.59 \\ 
    $S_0\to S_2 (\pi\pi^*)$ & 8.93 & 10.46 & 10.06 & 10.34 & 9.17 & 9.37 \\
    $S_2\to S_1$ & 2.73 & 2.75 & 0.89 & 1.07 & 0.82 & 0.78 \\
    \hline
  \end{tabular}
  \caption{Transitions energies in eV of the two lowest lying electronic states in methaniminium cation obtained with different methods$^\ddagger$.}
  \label{tbl:metAbs}
   \begin{tablenotes}
       \small
       \item $t) $ TD-DFTB/mio. $x) $ TD-PBE/6-31G* $a) $ CASSCF(12,8)/6-31G*. $b) $ FOMO-CASCI(12,8)/6-31G*($\beta$ = 0.2). $c) $ MR-CISD+Q/SA-9-[CAS(6,4)+AUX(4)]/d-aug-cc-pVDZ. $d) $ MRD-CI/6-31G**.
       \item $\ddagger $ All of these calculations were performed on their respective optimized structure in the ground state.
   \end{tablenotes}
\end{center}
\end{table}

\begin{table}[H]
\begin{center}
  \begin{tabular}{ c c c c c c }
    \hline
    Property & This work & Reference A\cite{barbatti2007fly} & Reference B\cite{barbatti2006ultrafast} & Reference C\cite{tapavicza2007trajectory} & Reference D\cite{du2015fly}\\
    \hline
    Method/level & (t) & (a) & (b) & (c) & (d) \\
    $\tau(S_2)$ [fs] & 14 & 12 & 13 & 10-50 & 8 \\ 
    $\tau(S_1)$ [fs] & 76 & 65 & 43 & 10-100 & 30-100 \\
    \# Trajs & 200 & 100 & 100 & 20 & 100 \\
    Total Time [fs] & 200 & 200 & 200 & 100 & 150 \\
    \hline
  \end{tabular}
  \caption{Excited states properties obtained from NAMD simulations using different methodologies and electronic structure methods.}
  \label{tbl:formNAMD}
   \begin{tablenotes}
       \small
       \item $t) $ TD-DFTB/mio. $a) $ MR-CISD/SA-3-CAS(4,3)/6-31G*. $b) $ SA-3-CAS(4,3)/6–31G*. $c) $ PBE/70 Ry Pseudopotentials. $d) $ B3LYP/6-31G**.
   \end{tablenotes}
\end{center}
\end{table}

\begin{figure}[H]
    \centering
    \includegraphics[width=1.0\linewidth]{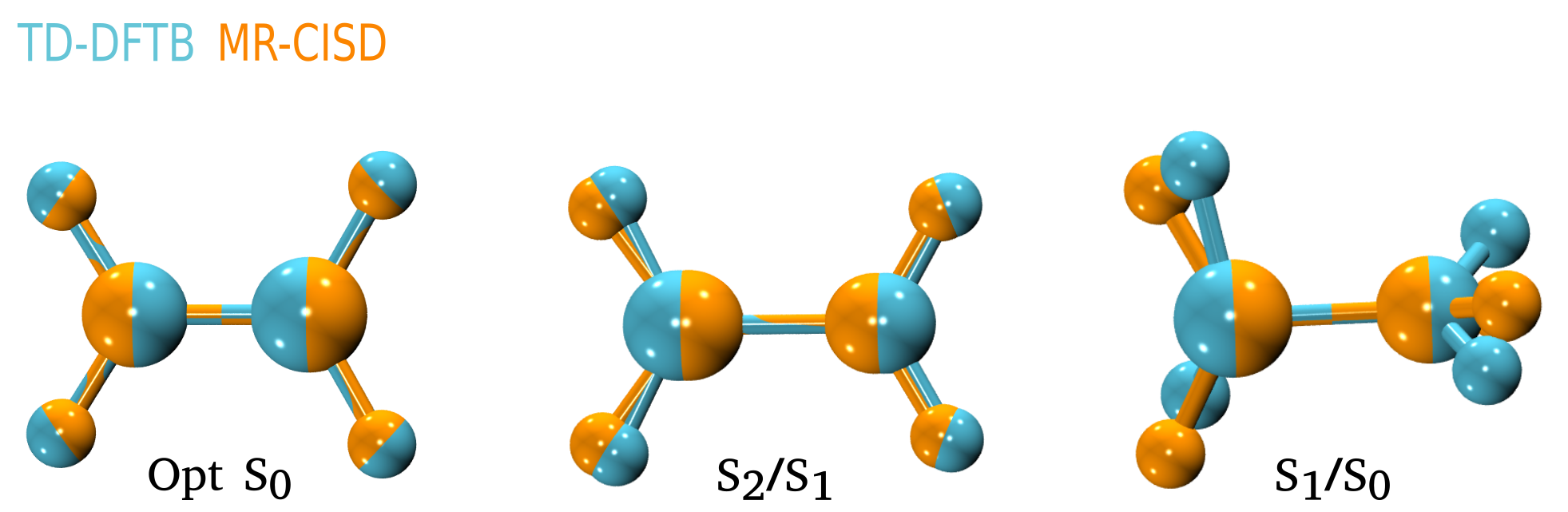}
    \caption{Superposition of the molecular geometries at ground state optimization (left panel), at the conical intersection (CoIn) $S_2/S_1$ (middle panel) and at the CoIn $S_1/S_0$ (right panel) obtained with DFTB/TD-DFTB in blue and with MR-CISD in orange (extracted from reference \citenum{barbatti2006ultrafast}).}
    \label{fig:formal}
\end{figure}

\section{Furan}

\begin{figure}[H]
    \centering
    \includegraphics{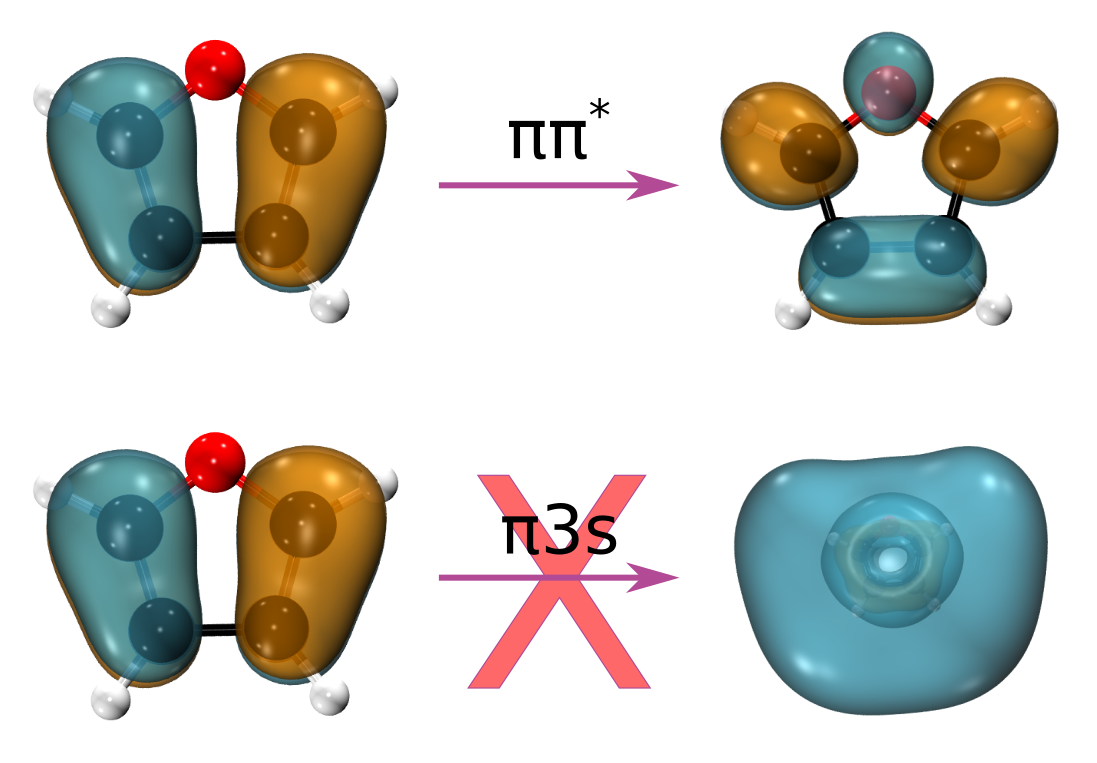}
    \caption{Molecular orbitals involved in both transitions in furan system. Rydberg state ($\pi 3s$) can not be calculated within TD-DFTB\cite{niehaus2021ground}.}
    \label{fig:mosfuran}
\end{figure}

\begin{table}[H]
\begin{center}
  \begin{tabular}{ c c c c c }
    \hline
    Property & This work & Reference A\cite{fuji2010ultrafast} & Reference B\cite{filatov2021signatures} & Reference C\cite{oesterling2017substituent} \\
    \hline 
    Method/level & $(t)$ & $(a)$ & $(b)$ & $(c)$ \\
    $\tau(S_0)$ [fs] & 84.5 & 60.0 & 68.0 & 95.0 \\
    \% $CoIn_{puck}$ & 70 & main & 86 & 10 \\
    \% $CoIn_{ropn}$ & 30 & low  & 14 & 90 \\
    Total Time [fs] & 200 & 200 & 250 & 250 \\
    \# Trajs & 200 & 240 & 100 & 100 \\
    \hline
  \end{tabular}
  \caption{Excited states properties obtained from NAMD simulations using different methodologies and electronic structure methods.}
  \label{tbl:furanNAMD}
   \begin{tablenotes}
       \small
       \item $t) $ TD-DFTB/mio. $a) $ TDDFT-PBE0/6-311++G**. $b) $ DISH-XF/SSR-BH\&HLYP/6-311G**. $c)  $ SA-CASSCF(10,9)/6-31G*
   \end{tablenotes}
\end{center}
\end{table}

\section{Molecular motors}

\begin{figure}[H]
    \centering
    \includegraphics{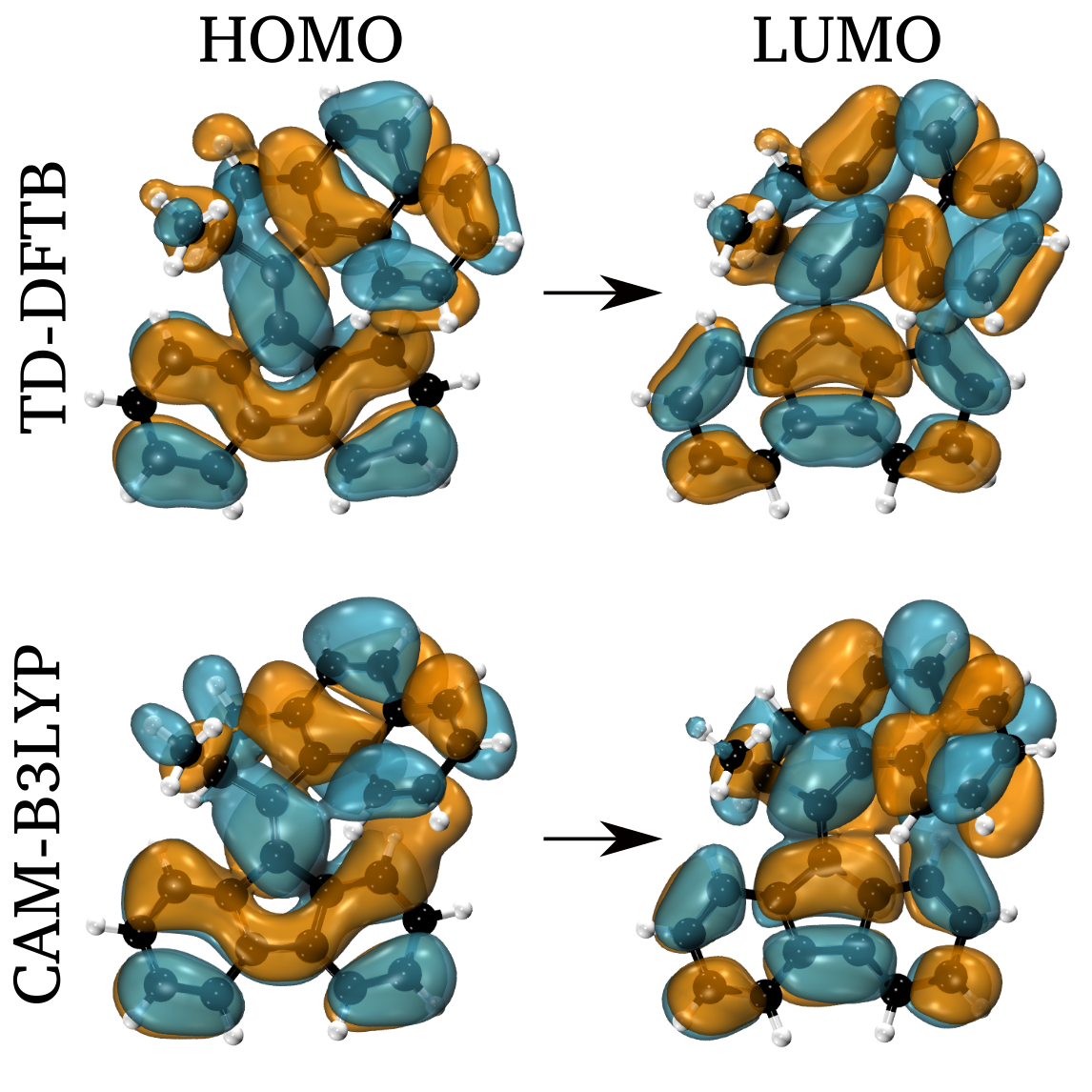}
    \caption{Molecular orbitals involved in the $S_0\to S_1$ transition at the Franck-Condon region obtined with TD-DFTB and with CAM-B3LYP/6-31G* in the molecular motor.}
    \label{fig:mosMotor}
\end{figure}

\begin{table}[H]
\begin{center}
  \begin{tabular}{ c c c c }
    \hline
    Excitation & TD-DFTB$^a$ & TD-LC-DFTB$^b$ & CAM-B3LYP$^c$ \\
    \hline 
    $S_0\to S_1$ [eV] & 2.6 (0.00) & 3.5 (0.56) & 3.6 (0.71) \\
    $S_0\to S_2$ [eV] & 2.7 (0.36) & 3.9 (0.01) & 3.8 (0.01) \\
    \hline
  \end{tabular}
  \caption{Excitation energies and oscillator strentgh (in brackets in a.u.) for the two lowest-lying transitions in molecular motors with different methods.}
  \label{tbl:MotorsMethod}
   \begin{tablenotes}
       \small
       \item $a) $ TD-DFTB/mio. $b) $ TD-DFTB with LC (used in this work). $c) $ TD-DFT/CAM-B3LYP/6-31G*. 
   \end{tablenotes}
\end{center}
\end{table}

\begin{figure}[H]
    \centering
    \includegraphics{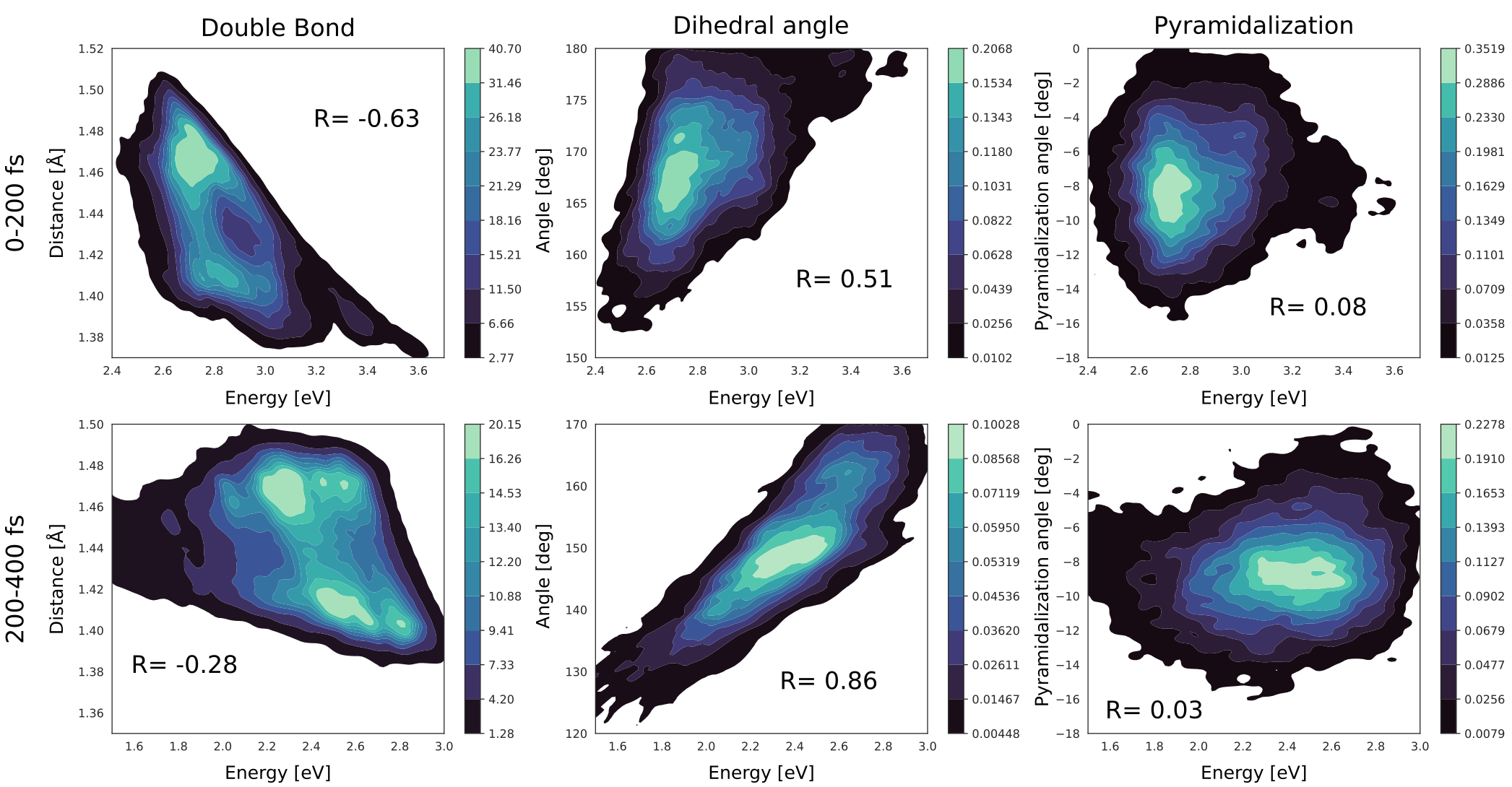}
    \caption{Waiting time analysis. 2D histograms of the first regime (0-200 fs, upper panels) and second regime (200-400 fs, lower panels) for the double bond (left panels), dihedral angle (middle panels) and pyramidalization angle (right panels) with the energy gap. The color intensity represents the density population. The $R$ value represent the Pearson correlation factor between the modes and the energy gap for both regimes.}
    \label{fig:corrEneMotors}
\end{figure}

\begin{figure}[H]
    \centering
    \includegraphics{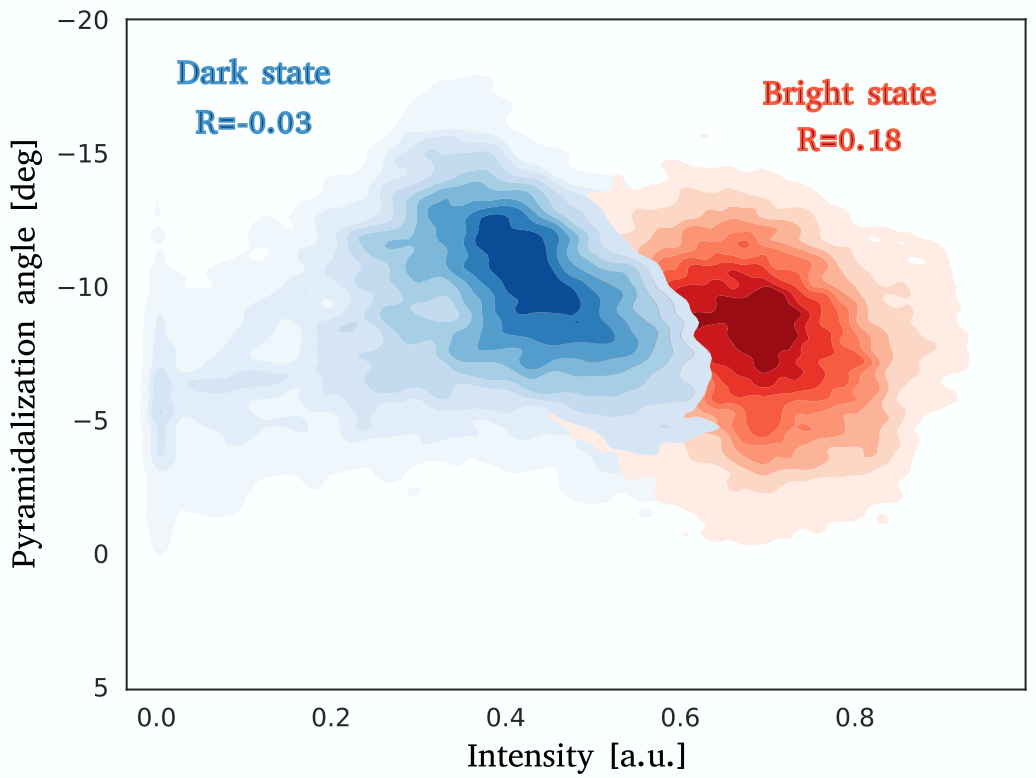}
    \caption{2D histogram of the distribution for the pyramidalization angle in both bright and dark states. The $R$ value represent the Pearson correlation factor between the modes and the emission intensity.}
    \label{fig:corrPyra}
\end{figure}

\begin{figure}[H]
    \centering
    \includegraphics{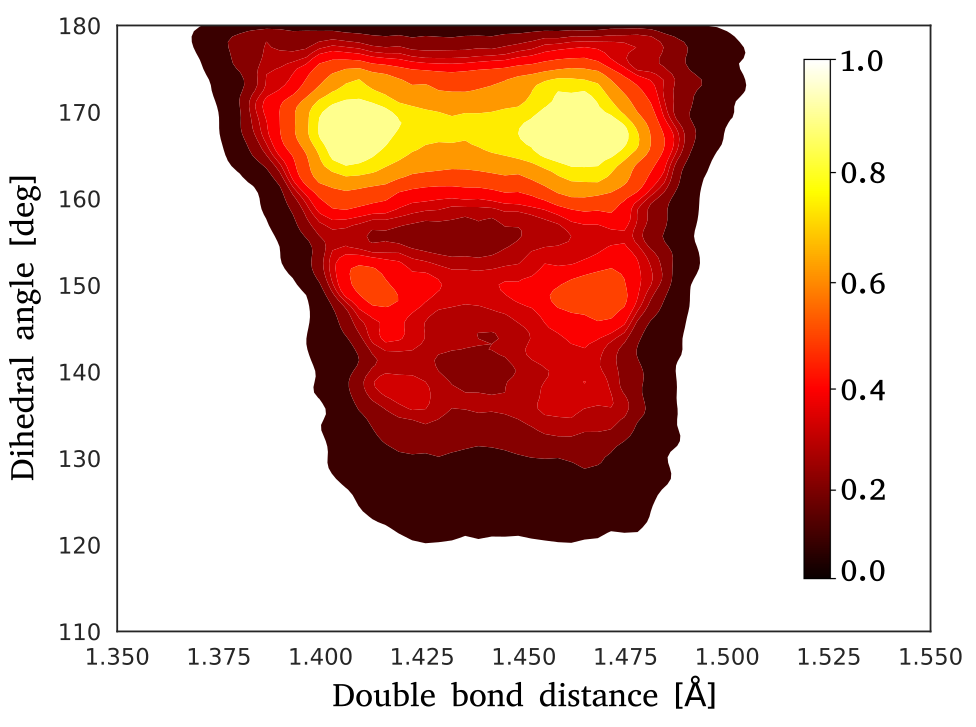}
    \caption{2D histogram of the distribution for the dihedral angle and double bond distances. The color scale shows the emission intensity in a.u.}
    \label{fig:2Dhist}
\end{figure}

\end{suppinfo}

\bibliography{bibref}